\documentclass[a4paper]{jpconf}
\usepackage{graphicx}
\usepackage{wrapfig}
\begin{document}
\title{Spin-Echo Measurements for an Anomalous Quantum Phase of 2D Helium-3}

\author{S. Takayoshi, K. Obata, D. Sato, T. Matsui and Hiroshi Fukuyama}

\address{Department of Physics, Graduate School of Science,
The University of Tokyo, 7-3-1 Hongo, Bunkyo-ku, Tokyo 113-0033, Japan}

\ead{takayoshi@phys.s.u-tokyo.ac.jp}

\begin{abstract}
Previous heat-capacity measurements of our group had shown
the possible existence of an anomalous
quantum phase containing the zero-point vacancies (ZPVs) in 2D $^{3}$He.
The system is monolayer $^{3}$He adsorbed on graphite preplated with
monolayer $^{4}$He at densities ($\rho$) just below the 4/7 commensurate phase
($0.8\leq \rho /\rho_{4/7}\leq 1$).
We carried out pulsed-NMR measurements in order to
examine the microscopic and dynamical nature of this phase.
The measured decay of spin echo signals shows the non-exponential behaviour.
The decay curve can be fitted with the double exponential function,
but the relative intensity of the component with a longer time constant
is small (5 \%) and does
not depend on density and temperature, which contradicts the macroscopic
fluid and 4/7 phase coexistence model.
This slowdown is likely due to the mosaic angle spread of Grafoil
substrate and the anisotropic spin-spin relaxation time $T_{2}$ in
2D systems with respect to the magnetic field direction.
The inverse $T_2$ value deduced from the major echo signal
with a shorter time constant,
which obeys the single exponential function, decreases
linearly with decreasing density from $n=1$, supporting the ZPV model.
\end{abstract}

\section{Introduction}
Monolayer $^{3}$He on Grafoil (exfoliated graphite) preplated with
$^{4}$He monolayer is an ideal model system
for strongly correlated two-dimensional (2D) fermions.
The advantage of this system is that the correlation can be varied in a
wide range by changing $^{3}$He areal density ($\rho$).
When $\rho$ is relatively small, the system behaves as a 2D Fermi fluid.
As $\rho$ increases, the correlation between the particles becomes stronger, 
and they localize at $\rho_{4/7}=6.80$ nm$^{-2}$ with the assistance
of the substrate potential corrugation \cite{Matsumoto_JLTP, Murakawa}.
This is attributed to the Mott-Hubbard transition \cite{Casey_PRL}.
The localized phase is a commensurate solid with the 4/7 density of
the first layer $^{4}$He, so called the 4/7 phase \cite{Elser_PRL}.
The 4/7 phase has a triangular lattice structure, which causes magnetic
fluctuation among the $^{3}$He nuclear spins.

Previous heat capacity measurements by our group strongly suggest the
existence of ``zero point vacancies (ZPVs)'' in a density range of
$0.8\leq n\equiv \rho /\rho_{4/7}\leq 1$ \cite{Matsumoto_JLTP}.
The ZPV is an atomic vacancy which exists stably even in the ground state.
It is spontaneously created in quantum solids when the half band-width
of quantum-mechanically hopping ZPVs exceeds the
vacancy creation energy.
Although the possible existence of ZPVs in solid He had been
proposed by Andreev and Lifshitz in 1969 \cite{Andreev_JETP},
it has not been found experimentally yet.
However, the ZPVs are supposed to emerge in 2D $^{3}$He system
because vacancies may be doped, retaining the 4/7 structure
in order to reduce the potential energy caused by the underlayer.

In this work, we carried out pulsed-NMR measurements with the spin-echo method which
reflect the microscopic and dynamical natures of the system.
If the macroscopic phase-separation happens rather than the emergence of
the ZPV phase, the NMR transverse relaxation should have two components,
each of which has the characteristic spin-spin relaxation time ($T_{2}$).
In other words, if the single exponential relaxation is observed
in the corresponding density region:
$0.8\leq n\leq 1.0$, the macroscopic phase separation does not happen.

\section{Experimental}
We used Grafoil substrate which consists of micro-crystallites (platelets)
with atomically flat surfaces of about 10 nm size \cite{Niimi_PRB}.
The mosaic angle spread of platelets is about $\pm$15$^{\circ}$ \cite{Taub_PRB}.
The total surface area is measured to be 53.6 m$^{2}$
from an N$_{2}$ adsorption isotherm at 77 K. The first layer ($^{4}$He)
is adsorbed at 4.2 K and the second layer ($^{3}$He) at 2.7 K.
The NMR transverse relaxation process was monitored by the spin-echo technique
with the pulse sequence of $90^{\circ}$-\,$t$\,-$180^{\circ}$-\,$t$
in a static magnetic field of 172 mT ($f=5.5$ MHz) parallel to the graphite basal plane.
We averaged the raw echo signals 50 to 2000 times depending on the signal amplitude.
Other experimental methods have been described before \cite{Murakawa}.

\section{Results and discussions}
\begin{figure}
\begin{minipage}{14pc}
\includegraphics[width=14pc]{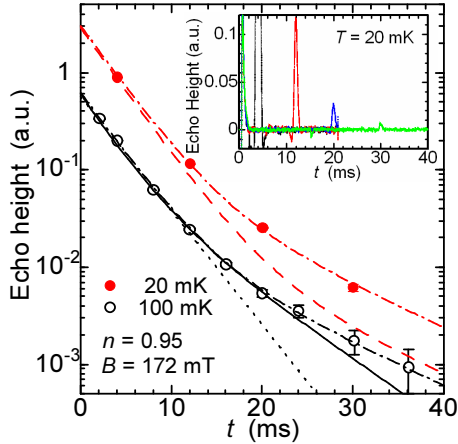}
\end{minipage}
\hspace{2pc}
\begin{minipage}{20pc}
\caption{\label{fig:n=0_95}
Spin echo height as a function of $t$
in the anomalous phase of 2D $^{3}$He at $n=0.95$.
$B=172$ mT. Closed and open circles
are data at $T=20$ and 100 mK, respectively. The dash-dotted lines are
the double exponential fitting (Eq.\,(\ref{eq:decay})).
The dotted line is the single exponential behavior
at $T=100$ mK representing only the first term in Eq.\,(\ref{eq:decay}).
The dashed line is the decay at 20 mK
estimated from the macroscopic two-phase coexistence model.
The solid line shows the decay at
100 mK calculated from Eq.\,(\ref{eq:angle}).
The inset shows raw echo signals.}
\end{minipage}
\end{figure}

\begin{wrapfigure}[17]{r}{16pc}
\begin{center}
\includegraphics[width=14pc]{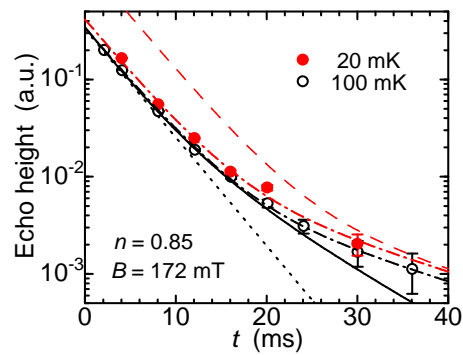}
\caption{\label{fig:n=0_85}
Spin echo height as a function of $t$
in the anomalous phase of 2D $^{3}$He at $n=0.85$.
Definitions for the symbols and lines are the same as Figure
\ref{fig:n=0_95}.}
\end{center}
\end{wrapfigure}

The measured transverse relaxations at $n=0.95$ and 0.85 are shown in Figures
\ref{fig:n=0_95} and \ref{fig:n=0_85}, respectively.
We carried out the measurements at $T=100$ and 20 mK because $T_{2}$ is independent
of $T$ in the temperature range of $10\leq T\leq 700$ mK
like the exchange plateau in bulk solid $^{3}$He \cite{Guyer_RMP}.
At first glance, the relaxations have two components.
The longer $T_{2}$ component is not due to the inappropriate background subtraction
(see the raw signals shown in the inset).

\begin{figure}[b]
\begin{minipage}{18pc}
\includegraphics[width=14pc]{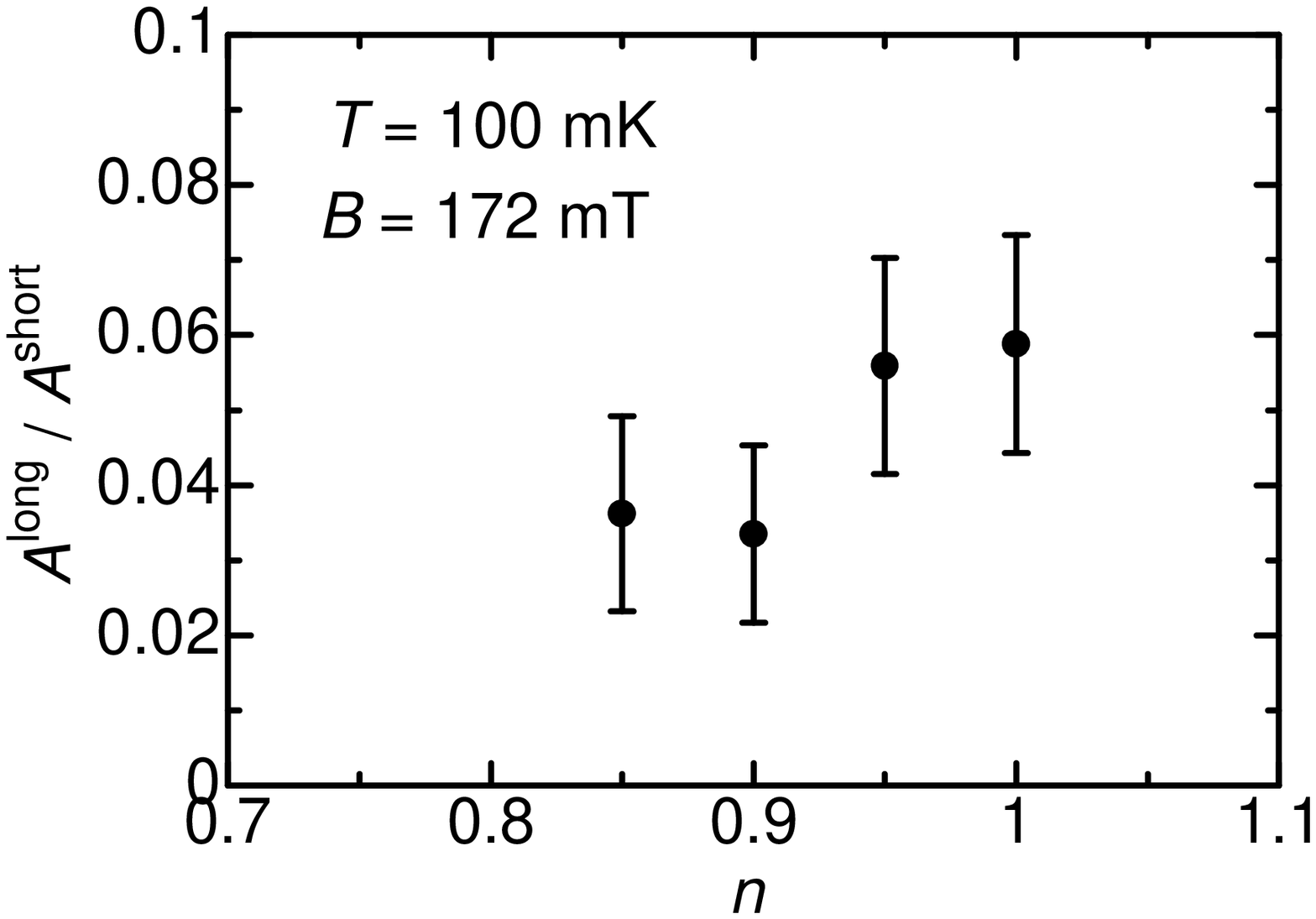}
\caption{\label{fig:prefactor}Density dependence of the ratio
$A^{\mathrm{long}}/A^{\mathrm{short}}$ of prefactors
in Eq.\,(\ref{eq:decay}).}
\end{minipage}
\hspace{2pc}
\begin{minipage}{18pc}
\includegraphics[width=13.6pc]{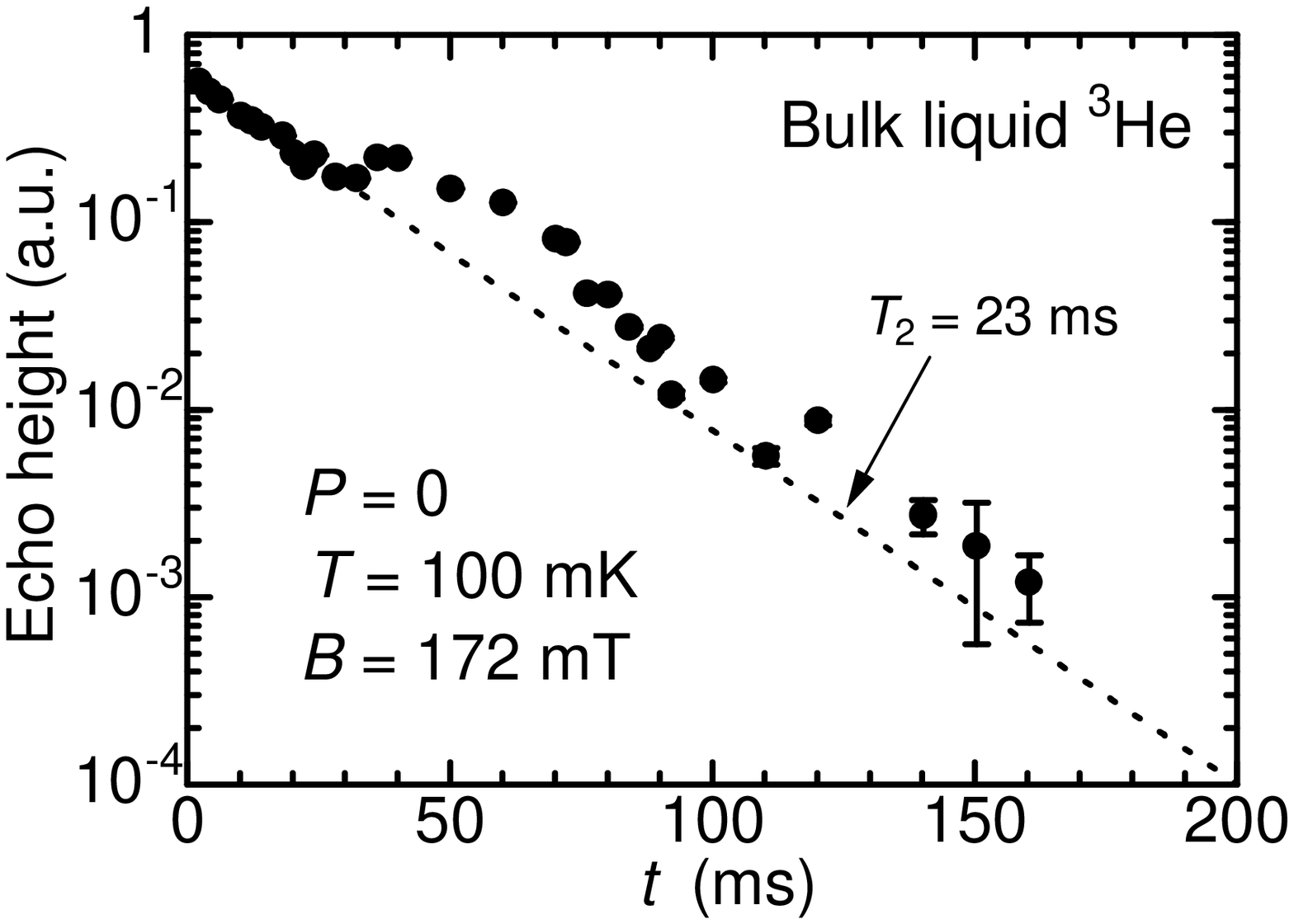}
\caption{\label{fig:bulk}Spin echo height as a function of $t$
in bulk $^{3}$He of 0 Pa at $T=100$ mK.
The dotted line is the extrapolation of the single exponential
behavior at $t\leq 30$ ms.}
\end{minipage} 
\end{figure}

Within the macroscopic two-phase coexistence model, the decay of the echo amplitude $S$
should follow a double exponential function
\begin{equation}
S=A^{\mathrm{short}}\exp (-t/T_{2}^{\mathrm{short}})
+A^{\mathrm{long}}\exp (-t/T_{2}^{\mathrm{long}}).
\label{eq:decay}
\end{equation}
The $T_{2}^{\mathrm{short}}$ and $T_{2}^{\mathrm{long}}$ components are
signals from the 4/7 phase and the high density Fermi fluid, respectively.
The results of fitting to Eq.\,(\ref{eq:decay}) are
the dash-dotted lines in Figures \ref{fig:n=0_95}
and \ref{fig:n=0_85}. Although the fitting quality is seemingly well,
this model is inadequate in several respects.
Firstly, in this model, the decay at $T=20$ mK is estimated from
the $T=100$ mK data as the dashed lines.
We used here the known $T$-dependence of magnetization for
the high-density Fermi fluid
and the 4/7 phase measured in Ref.\,\cite{Murakawa}.
These estimations do not describe the 20 mK data at all.
Secondly, the ratio $A^{\mathrm{long}}/A^{\mathrm{short}}$
remains unchanged or even decreases with decreasing $n$
as shown in Figure \ref{fig:prefactor}.
If the system consists of two components,
$A^{\mathrm{long}}/A^{\mathrm{short}}$ should increase linearly
with decreasing density in the coexistence region.
Therefore, the macroscopic two-phase coexistence model is clearly excluded.

We also made the same measurement filling the sample cell with
liquid $^{3}$He at $P=0$. 
The substrate surface is still preplated by a $^{4}$He monolayer.
As shown in Figure \ref{fig:bulk},
the slowdown of decay (the existence of slow component) is not observed in this case.
This means that the long $T_{2}$ component is characteristic
of the 2D samples.
The detailed structure in the decay around $t=30$ ms is probably
due to the complicated diffusion process
since the sample occupies spaces within and without the Grafoil stack.

Then, what is the slowdown of decay caused by?
Cowan \cite{Cowan_JPC} showed theoretically that $T_{2}$ depends strongly
on the angle ($\beta$) between the static magnetic field direction and
the vector normal to plane in 2D systems.
This effect has been confirmed by the previous
NMR experiments at $1.2\leq T\leq 4.2$ K \cite{Satoh_JLTP}.
If we assume that a Gaussian distribution
with a standard deviation $\sigma =15^{\circ}$
for the mosaic angle spread of Grafoil substrate,
$S$ is calculated by
\begin{equation}
S=\int_{-90^{\circ}}^{90^{\circ}}\mathrm{d}\beta
\exp \{-tT_{2}^{-1}(\beta)\}\frac{1}{\sqrt{2\pi}\sigma}
\exp\left\{-\frac{(\beta -90^{\circ})^{2}}{2\sigma^{2}}\right\} .
\label{eq:angle}
\end{equation}
We used here the $\beta$-dependence of $T_{2}$, $T_{2}(\beta)$,
for $\omega_{0}\tau_{0}=10^{-1}$ given in
Figure 4 of Ref.\,\cite{Cowan_JPC}, where $\omega_{0}= 2\pi f$ is
the Larmor frequency. $\tau_{0}$ is defined as $a^{2}/2D$
in which $a(= 4.1\times 10^{-10}\;\mathrm{m})$ is the lattice constant and
$D(\sim 10^{-11}\;\mathrm{m}^{2}\cdot\mathrm{s}^{-1})$ is the spin diffusion constant.
The fitted result is the solid lines in Figures
\ref{fig:n=0_95} and \ref{fig:n=0_85}.
Although the fitting quality is not very good because the mosaic angle distribution
of our substrate and the value of $D$ are not accurately known,
the anisotropy of $T_{2}$ explains the measured slowdown of decay semi-qualitatively.
The insensitivity of $A^{\mathrm{long}}/A^{\mathrm{short}}$ to $n$ and
$T$ can be naturally understood along this consideration
that the long tail component originates from the extrinsic effect
due to the substrate.

\begin{figure}[t]
\begin{minipage}{18pc}
\includegraphics[width=14pc]{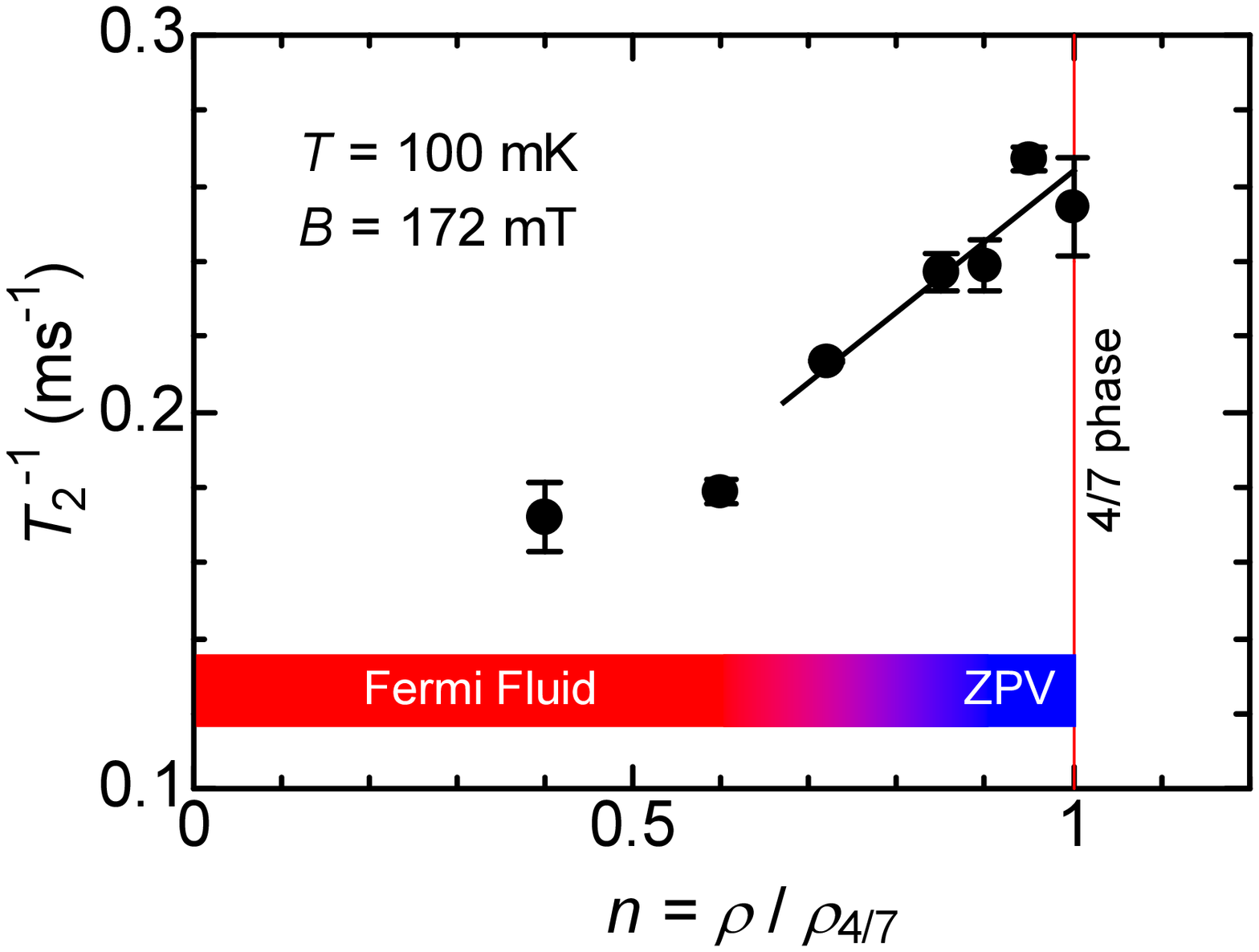}
\caption{\label{fig:dens_dep}Density dependence of $T_{2}^{-1}$ at $T=100$ mK.
The solid line is a guide to the eye.}
\end{minipage}
\hspace{2pc}
\begin{minipage}{18pc}
\includegraphics[width=14pc]{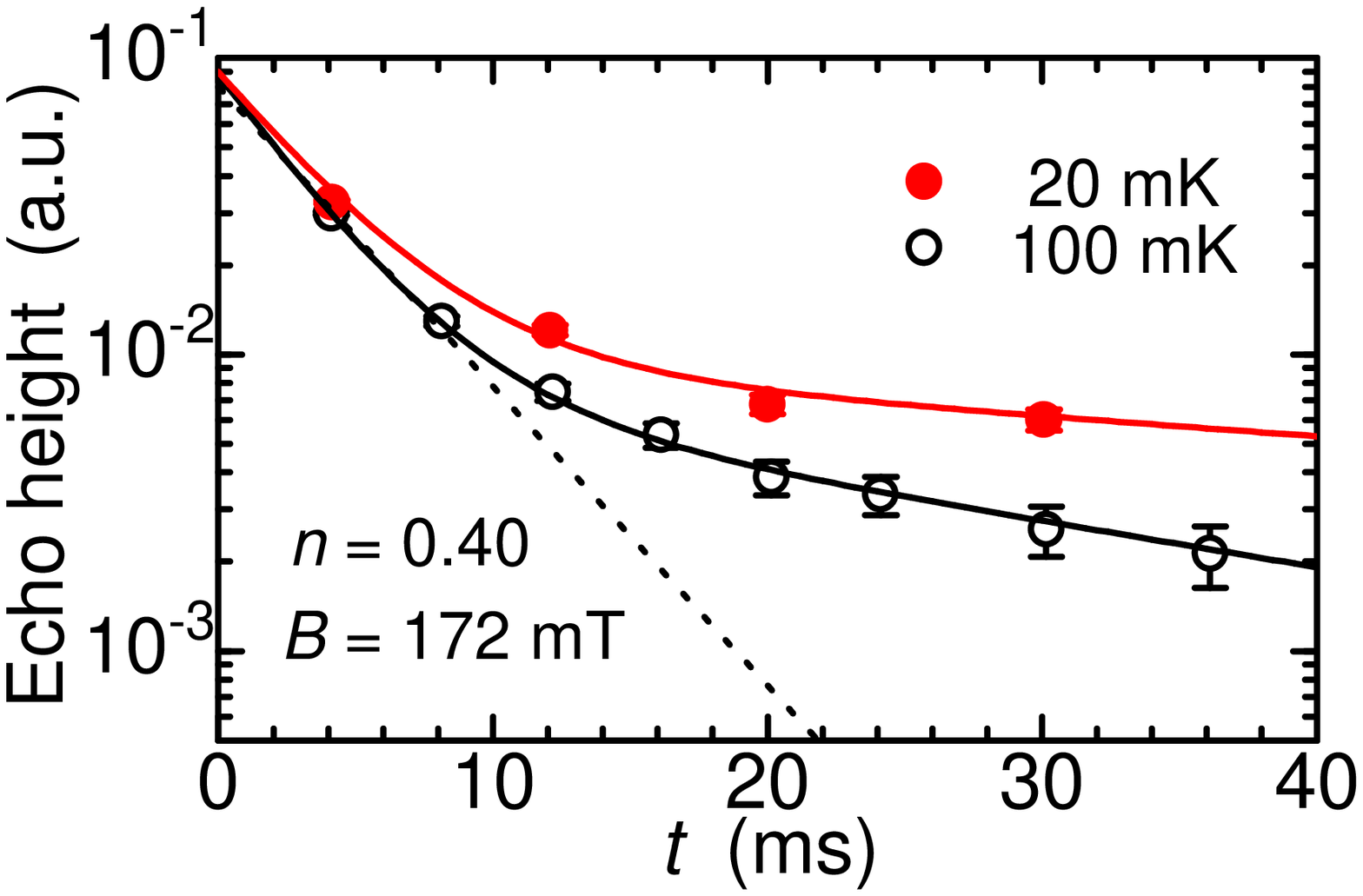}
\caption{\label{fig:n=0_40}Spin echo height as a function of $t$
in the Fermi fluid phase of 2D $^{3}$He at $n=0.40$.
The closed and open circles are data at $T=20$ and
100 mK, respectively. The solid lines are guides to the eye.}
\end{minipage}
\end{figure}

Therefore, the short $T_{2}$ component,
which contributes to the total magnetization by about 95 \%
(see Figure \ref{fig:prefactor}), is an intrinsic transverse relaxation,
i.e., $T_{2}^{\mathrm{short}}=T_{2}$.
In other words, the macroscopic phase separation model can be excluded.
The density dependence of $T_{2}^{-1}$ at $T=100$ mK
determined in this way is shown in Figure \ref{fig:dens_dep}.
In the region of $0.7\leq n\leq 1$,
$T_{2}^{-1}$ decreases monotonically with decreasing $n$.
This is consistent with the expectation from the ZPV model that the number of ZPVs
doped into the 4/7 phase linearly increases with decreasing $n$.

It should be noted that if the system is phase separated microscopically
on the length scale much shorter than the diffusion length $l_{D}$
within the time interval $T_{2}$, the fluid and 4/7 phase domains are
in the fast-exchange limit \cite{Hammel} and a single $T_{2}$ will be observed.
We estimate $l_{D} \sim 200$ nm from the relation $l_D \sim a \sqrt{T_{2}J}$
where $J$ is the exchange frequency. Thus, our experimental results
do not exclude the possibilities of the microscopic phase separation
on the length scale shorter than
several tens nm due to either intrinsic (e.g. domain wall structures)
or extrinsic (e.g. substrate heterogeneities \cite{Morhard}) effects.

Finally, we briefly discuss the data obtained in the Fermi fluid phase
($n=0.40$) shown in Figure \ref{fig:n=0_40}.
The echo signal extrapolated to $t=0$, i.e. the magnetization,
is unchanged between $T=20$ and 100 mK,
which is characteristic of degenerated Fermi fluid.
The decay rate also decreases with increasing $t$ here.
However, this does not originate only from the $T_{2}$ anisotropy,
but could be related to the Fermi liquid effects such as increasing $D$
with decreasing $T$, because both $T_{2}^{\mathrm{short}}$ and $T_{2}^{\mathrm{long}}$
vary with $T$.
Further measurements of the detailed $T$-dependence of the relaxation and
of $D$ values under field gradients will provide us useful
information on the spin diffusion in 2D fermions.

This work was financially supported by Grant-in-Aid for Scientific
Research on Priority Areas (No. 17071002) from MEXT, Japan.

\end{document}